\documentclass{aastex}
\usepackage{emulateapj5}

\begin{document}

\shortauthors{Gordon et al.\ 2000}
\shorttitle{Dust Emission Features in NGC 7023}

\slugcomment{accepted on 10 Jul 2000 for publication in the ApJ}

\title{Dust Emission Features in NGC 7023 between 0.35 and 2.5 $\micron$: 
Extended Red Emission (0.7~$\micron$) and 
Two New Emission Features (1.15 and 1.5~$\micron$)} 

\author{Karl D.\ Gordon,\altaffilmark{1}
   Adolf N.\ Witt\altaffilmark{2}, 
   Richard J.\ Rudy\altaffilmark{3}, 
   R.\ C.\ Puetter\altaffilmark{4},
   David K.\ Lynch\altaffilmark{3},
   S.\ Mazuk\altaffilmark{3},
   K.\ A.\ Misselt\altaffilmark{5,6},
   Geoffrey C.\ Clayton\altaffilmark{5,6}, \&
   Tracy L.\ Smith\altaffilmark{2}}
\altaffiltext{1}{Steward Observatory, University of Arizona,
   Tucson, AZ 85721; kgordon@as.arizona.edu}
\altaffiltext{2}{Ritter Astrophysical Research Center, The University
   of Toledo, Toledo, OH 43606; awitt@dusty.astro.utoledo.edu; 
   tsmith@astro1.astro.utoledo.edu}
\altaffiltext{3}{Space Science Applications Laboratory, M2/266, 
   The Aerospace Corporation, P.O.\ Box 92957, Los Angeles, CA 90009; 
   (richard.j.rudy,david.k.lynch,steve.mazuk)@aero.org}
\altaffiltext{4}{Center for Astrophysics and Space Sciences, C-0111, 
   University of California, San Diego, La Jolla, CA  92093; 
   rpuetter@ucsd.edu}     
\altaffiltext{5}{Department of Physics \& Astronomy, Louisiana State
   University, Baton Rouge, LA 70803; (gclayton,misselt)@fenway.phys.lsu.edu}
\altaffiltext{6}{Visiting Astronomer, Kitt Peak National Observatory,
   National Optical Astronomy Observatories, which is operated by the
   Association of Universities for Research in Astronomy, Inc.\ (AURA)
   under cooperative agreement with the National Science Foundation. }

\begin{abstract}
We present 0.35 to 2.5~$\micron$ spectra of the south and northwest
filaments in the reflection nebula NGC 7023.  These spectra were used
to test the theory of Seahra \& Duley that carbon nanoparticles are
responsible for Extended Red Emission (ERE).  Our spectra fail to show
their predicted second emission band at 1.0~$\micron$ even though both
filaments exhibit strong emission in the familiar 0.7~$\micron$ ERE
band.  The northwest filament spectrum does show one, and possibly
two, new dust emission features in the near-infrared.  We clearly
detect a strong emission band at 1.5~$\micron$ which we tentatively
attribute to $\beta$-FeSi$_2$ grains.  We tentatively detect a weaker
emission band 
at 1.15~$\micron$ which coincides with the location expected for
transitions from the conduction band to mid-gap defect states of silicon
nanoparticles.  This is added evidence that silicon nanoparticles are
responsible for ERE as they already can explain the observed behavior
of the main visible ERE band.
\end{abstract}

\keywords{infrared: ISM: lines and bands -- ISM: individual (NGC 7023)
-- ISM: lines and bands -- reflection nebulae }

\section{Introduction \label{sec_intro}}

Recently, \citet{sea99} proposed carbon nanoparticles as the material
responsible for Extended Red Emission (ERE).  ERE is a broad
($\Delta\lambda \sim 0.1~\micron$) emission band with a peak
wavelength between 0.65 and 0.88~$\micron$ seen in many
dusty astrophysical objects as well as the diffuse interstellar medium
(\citet{gor98} and references therein).  A review of the observed
characteristics of ERE can be found in \citet{wit98}.  Among other
proposed carriers for ERE are hydrogenated amorphous carbon
\citep{dul85}, coal \citep{pap96}, quenched carbonaceous composite
\citep{sak92}, $C_{60}$ \citep{web93}, and silicon nanoparticles
\citep{led98, wit98, led00}.  Identifying the material responsible for ERE is
important as \citet{gor98} have shown that this material must be a
major component of dust grains.  They found that the ERE in the
diffuse interstellar medium on a galaxy-wide scale emits 4\% of the
energy absorbed by dust at wavelengths below 0.55~$\micron$.  If ERE
could be convincingly identified with carbon nanoparticles, this could
have implications on the carriers of the 2175~\AA\ extinction bump and
3.4~$\micron$ CH$_n$ absorption feature as both have also been
attributed to carbon nanoparticles \citep{her98, sch98, sch99, dul99,
sea99}.

\citet{sea99} modeled the photoluminescence spectrum expected for
different sized carbon nanoparticles.  The carbon nanoparticles in
their model are composed exclusively of carbon and hydrogen atoms with
$sp^2$ bonds (i.e., aromatic bonds).  In general, the term carbon
nanoparticles refers to materials which a mixture of $sp^2$ and $sp^3$
bonds (i.e., aromatic and aliphatic bonds).  \citet{sea99}
specifically chose to model only $sp^2$ carbon nanoparticles to match
the observed high photoluminescence efficiency of ERE in the diffuse
ISM \citep{gor98}.  The \citet{sea99} model predicts an ERE
spectrum with three peaks at 0.5, 0.7, and 1.0~$\micron$.  The
0.7~$\micron$ peak is the peak usually identified with ERE.  The
0.5~$\micron$ peak has never been seen and this fact is accounted for
by \citet{sea99}.  In their model, the 0.5~$\micron$ peak is produced
by the smallest carbon particles.  They point out that the
environments where good ERE spectra exist are those with harsh UV
radiation fields which will destroy all carbon nanoparticles with less
than 50 carbon atoms.  This effectively removes the 0.5~$\micron$ peak
from the emission spectrum of ERE.  There is an observed ERE spectrum
of a high latitude cirrus cloud where the 0.5~$\micron$ sideband might
be expected, but it is not of sufficient signal-to-noise to detect the
0.5~$\micron$ peak \citep{szo98}.  The search for the predicted
1.0~$\micron$ band was the motivation for this work.  Previous to this
paper, no published 1.0~$\micron$ spectra existed of a dusty object
showing the familiar 0.7~$\micron$ ERE feature.

In order to test the prediction of a 1.0~$\micron$ feature associated
with ERE emitting dust, we have taken 0.35 to 2.5~$\micron$ spectra of
two filaments in NGC 7023.  NGC 7023 is the well known reflection
nebula illuminated by the Herbig Be star HD 200775
\citep{wit80, wit92, sel92, lem96}.  The two filaments we observed are
well known photodissociation regions which have been studied
extensively at many wavelengths (\citet{fue00} and references
therein).  These filaments are the tips of fingers of molecular
material, as seen in CO, which point towards HD 200775 \citep{ger98}.
The filaments have been observed to emit ERE \citep{wit90}, many $H_2$
emission lines \citep{lem96, lem99, mar97, mar99}, continuum emission
from transitionally heated small dust particles \citep{sel83, sel92},
and the Aromatic Infrared Features \citep{sel85, ces96, mou99, uch00}.
The two filaments are composed of molecular cloud material which is
being uncovered and modified by the UV-strong spectrum of HD 200775.
This results in a rich spectrum of dust and gas emission features.  In
addition to the dust and molecular emission features listed above, the
filaments are also known emitters of atomic hydrogen, carbon, and
oxygen \citep{fue00}.  Interestingly, the emission from atomic
silicon (\ion{Si}{2} 34.8~$\micron$) apears strongly enhanced between
the filaments and the illuminating star \citep{fue00}. 
This implies that there is a transition from the solid to the gas
phase of silicon at the filament edge.

\section{Observations \label{sec_obs}}

Long--slit optical spectra of NGC~7023 were obtained at KPNO with the
GoldCam spectrometer on the 2.1m telescope on 27 June 1998 in
photometric conditions.  Observations were carried out using grating
\#201 with a slit width of 3${\arcsec}$, resulting in a spectral range
of $0.35-1.0$~\micron\ with a resolution of $18$~\AA\ (FWHM)
as measured from night sky lines.  Two five second exposures of HD
200775 were obtained along with three 900 second exposures of NGC
7023; both sets of observations were obtained at an airmass of $1.3$.
For the observations of NGC 7023, the 5$\arcmin$ long slit was 
positioned $20\arcsec$ west of HD 200775 at a position angle of
$5\fdg 7$.  Standard reduction steps were performed using IRAF, and
the details of the reduction are given in \citet{mis99}.  The HD
200775 spectra were extracted from a $11\farcs 7 \times 3\arcsec$ wide
region centered on the star.  The sky emission was subtracted using
sky spectra determined from positions above and below the star.
Nebular spectra were extracted for the northwest and south filaments
and their sky subtraction was done using sky spectra measured near the
ends of the slit.

\begin{figure*}[tbp]
\epsscale{2.0}
\plotone{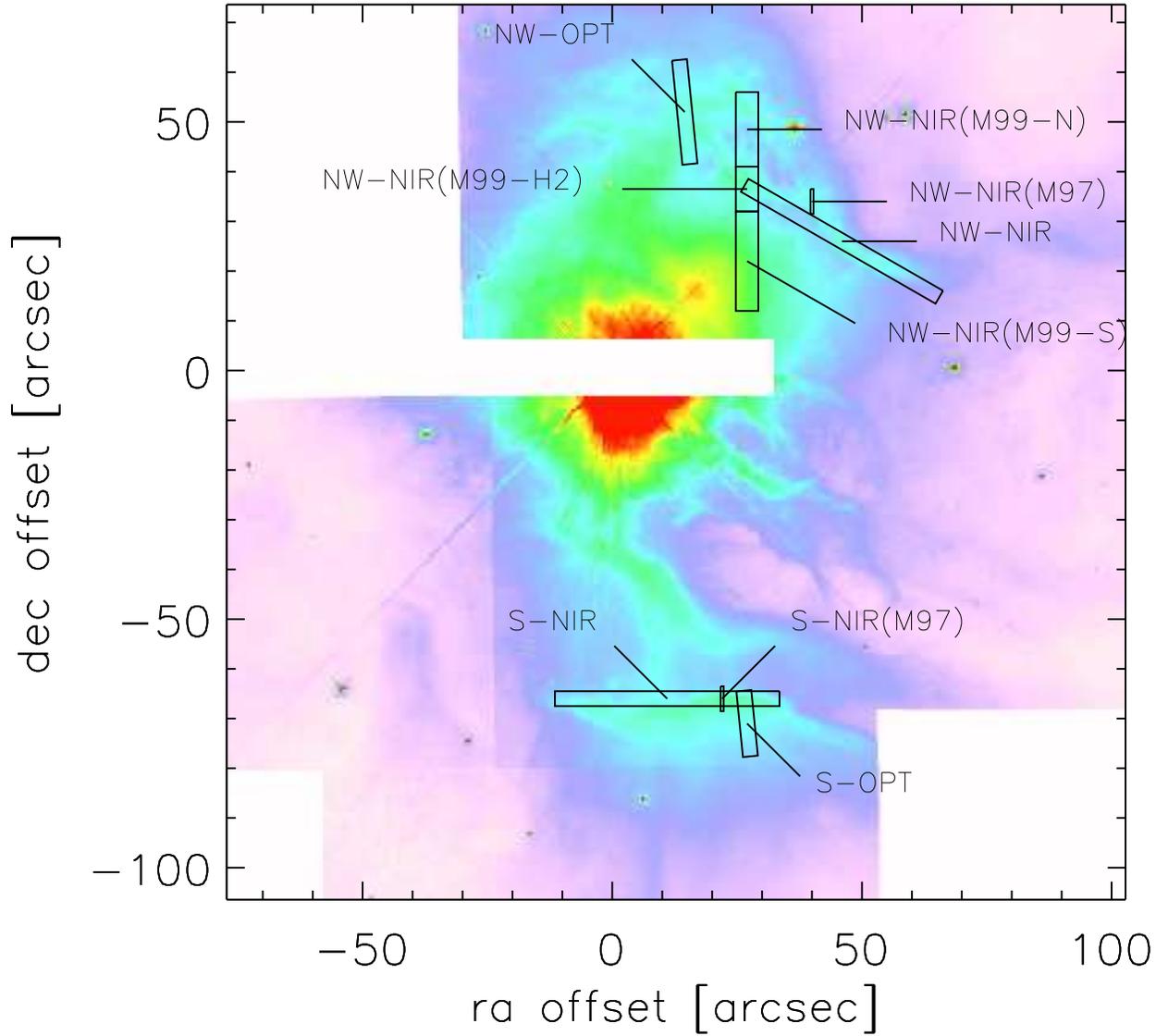}
\caption{The positions of optical and near-IR spectra are shown on a
WFPC2/F606W image of NGC 7023 constructed from HST archival
observations (\#5983, \citet{sta97}).  Eight individual F606W images
were used to construct the image, four at each of two positions, with
a range of exposure times.  The total exposure time for final image
was 1905 seconds.  The positions, sizes, and position angles for all
the spectra are given in Table~\ref{tab_pos}. \label{fig_pos}}
\end{figure*}

The positions, aperture sizes, and position angles of the northwest
and south filament spectra used in this paper are given in
Table~\ref{tab_pos}.  The optical spectra are designated NW-OPT and
S-OPT for the northwest and south filaments, respectively.  The other
entries are for the new near-infrared spectra presented in this paper
(NW-NIR and S-NIR) and previously published near-infrared spectra
\citep{mar97,mar99}.  Figure~\ref{fig_pos} shows the positions of
all the filament spectra on a WFPC2/F606W image constructed from HST
archival observations \citep{sta97}.

Our near-infrared spectrophotometry of the NGC 7023 filaments and its
illuminating star HD 200775 were acquired with the 3-m Shane reflector
of Lick Observatory.  The northwest and south filaments (NW-NIR and
S-NIR in Table~\ref{tab_pos}) were observed on 30 August 1999 (UT) and
HD 200775 was observed on both 28 and 30 August 1999 (UT).  The
instrument used was the Near InfraRed Imaging Spectrograph (NIRIS).  A
detailed description of the instrument, which is a two-channel, long
slit spectrograph operating from 0.8 to 2.5 $\micron$, is given by
\citet{rud99}.  The spectrograph has a beam splitter at 1.4
$\micron$ which feeds two arrays, allowing for simultaneous
observations below and above 1.4 $\micron$.  The arrays do not cover
the entire spectral range in one setting, but each requires 3 settings
to cover (with overlaps) the entire 0.8 to 2.5 $\micron$ range.  The
resolution of the spectra is 18 and 36~\AA\ for wavelengths shorter
and longer than 1.4~$\micron$, respectively.

For the observations of the south filament, the brightest portion of
the filament was centered visually in one half of the east/west
oriented slit.  For the northwest filament, the brightest portion of
the filament was centered visually in one half of a
northwest/southeast (position angle $= 60^{\circ}$) oriented slit.  To
facilitate removal of the telluric background emission, spectra were
acquired with the filament positioned in one half (position 1) and
then the other half (position 2) of the slit.  At each spectral
setting, four spectra, with exposure times of 25.5 seconds, were taken
in the pattern of position 1, 2, 2, and 1.  The total exposure time
for the filament spectra were 102 seconds.  The dimensions of the slit
were $3\arcsec \times 120\arcsec$.  The data from a $45\arcsec$ swath
along the slit at 
each of the two positions were combined to produce the filament
spectrum.  For the observations of HD 200775, a $20\arcsec$ telescope
``nod'' was used to separate the two observing positions.  The
individual exposure times for HD 200775 were 0.5 seconds, and 20
exposures were taken at each spectral setting.  The final spectrum of
HD 200775 is a combination of the data taken on 28 and 30 August 1999.

The comparison star used to remove the instrumental response and the
effects of atmospheric absorption, and to determine the absolute flux
level for both the filaments and their illuminating star on 30 August
1999 was HR 7967, a G8III star.  The visual magnitude of $V = 6.41$
from of the Bright Star Catalog \citep{hof82} and the $V-K$ color for
a G8 giant \citep{koo83}, implies a K magnitude of 4.25.  The
intrinsic spectral shape of HR 7967 was removed from this ratio by
using the model for this spectral type from \citet{kur91}.  On 28
August 1999, the comparison star was 16 Cyg B (HR 7504), a G3V star.
The positions, aperture sizes, and position angles of these
near-infrared filament spectra are given in Table~\ref{tab_pos} and
plotted in Figure~\ref{fig_pos}.

\begin{figure*}[tbp]
\epsscale{1.5}
\plotone{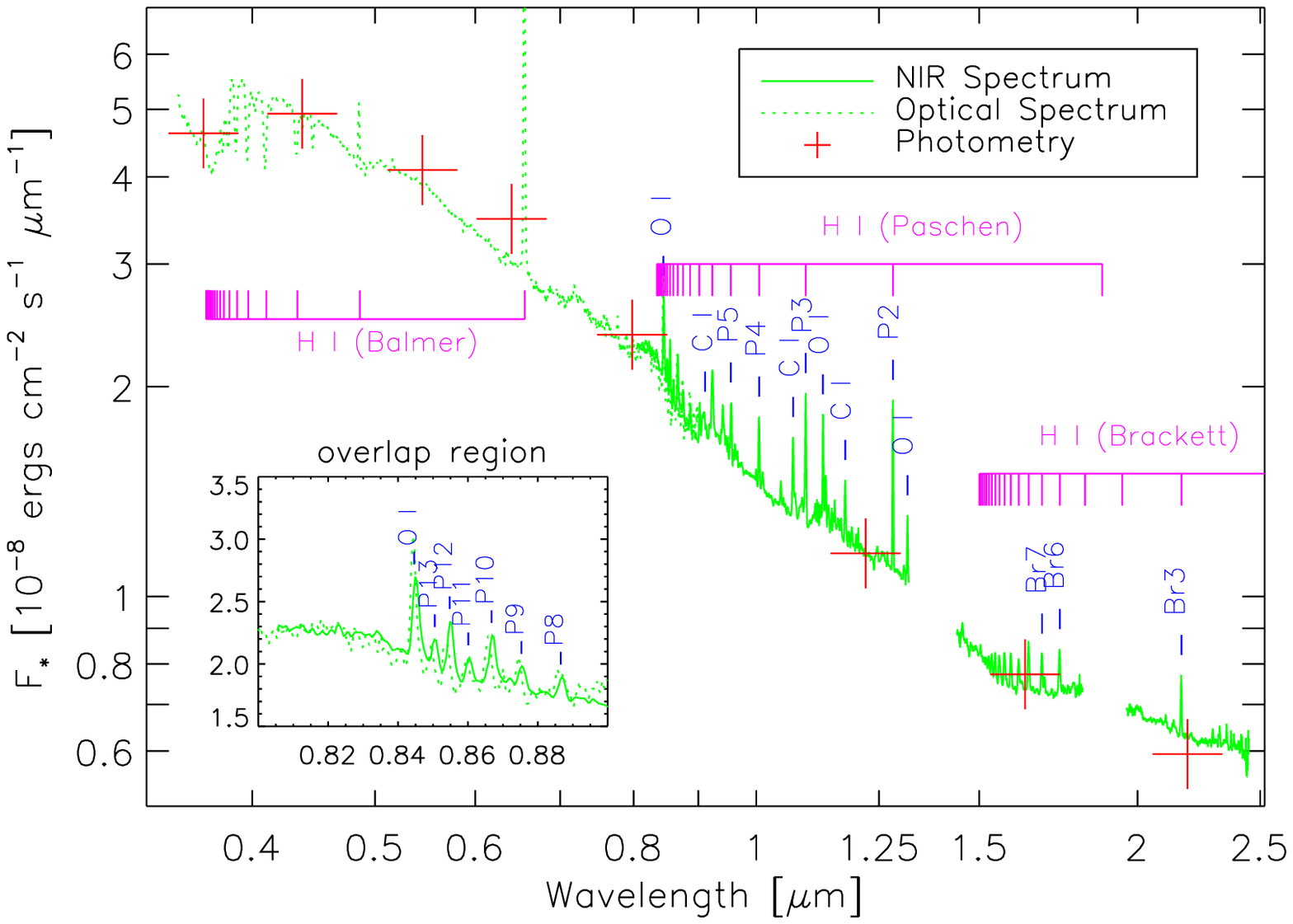}
\caption{The full optical and near-infrared spectrum of HD 200775 is
plotted.  Photometry from various sources is plotted as crosses.  The
Balmer, Paschen, and Brackett \ion{H}{1} lines are labeled as well as
the major \ion{O}{1} and \ion{C}{1} lines.  The inset plot gives a
closeup of the overlap region illustrating the quality of the relative
calibration of the optical and near-infrared observations.
\label{fig_star}}   
\end{figure*}

The combined 0.35--2.5~$\micron$ spectrum of HD 200775, the
illuminating star of NGC 7023, is plotted in Figure~\ref{fig_star}.
In addition, optical and near-infrared photometry for HD 200775 are
plotted \citep{all73, sel83, wit88, sel96, her99}.  The optical
spectrum was multiplied by 1.05 to make it match the optical
photometry and the overlap region with the near-IR spectroscopy.  This
adjustment could be due to the uncertainties in the absolute
calibration of the optical spectroscopy or the known variation in the
flux levels for HD 200775 \citep{her99}, a Herbig Be star.  The good
agreement between the optical and near-IR spectra in the overlap
region (Fig.~\ref{fig_star}, inset plot) in both continuum shape and
line features increases our confidence in both observations.

\begin{figure*}[tbp]
\epsscale{2.3}
\plottwo{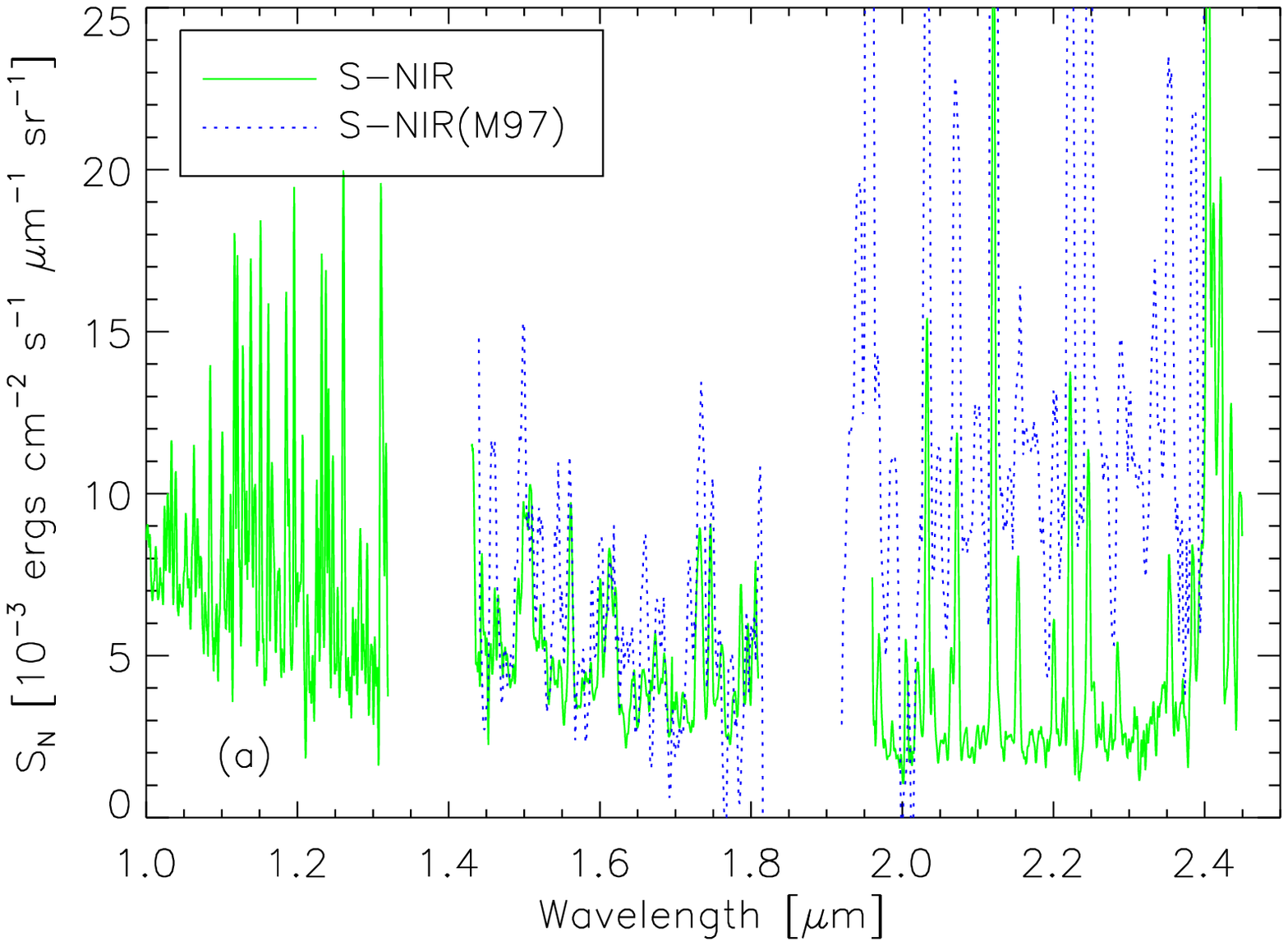}{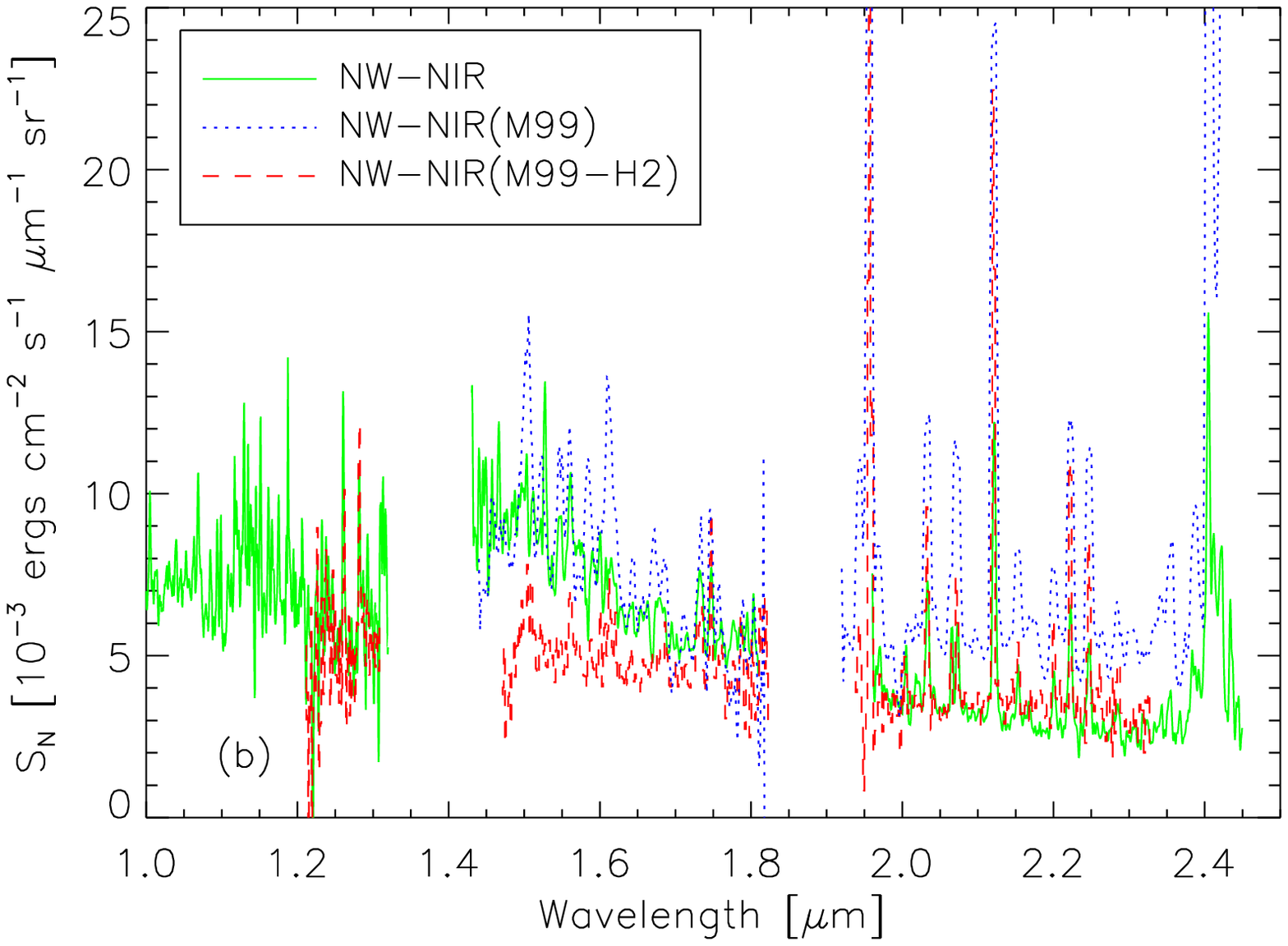}
\caption{The near-infrared spectra for the (a) south and (b) northwest
filaments are shown for the \citet{mar97} (S-NIR(M97) \& NW-NIR(M97),
H and K band only), the \citet{mar99} (NW-NIR(M99-H2), J, H, and K
band), and our new observations (S-NIR \& NW-NIR).  The \citet{mar97}
spectra have been smoothed with a 5 pixel wide boxcar and have been
multiplied by factors of 10 and 2.5 for (a) and (b), respectively, to
ease the spectral shape comparisons.  The \citet{mar99} spectrum (b
only) has been multiplied by 0.5 for the same reason.  The noise-like
appearance of the near-infrared spectra is due to a virtual forest of
$H_2$ emission lines.
\label{fig_comp_neb}} 
\end{figure*}

Our new near-infrared spectra of the south and northwest filaments are
plotted along with those from the literature \citep{mar97, mar99} in
Figure~\ref{fig_comp_neb}.  As the literature observations were taken
with different apertures and on similar, but different, locations in
each filament (see Table~\ref{tab_pos} and Fig.~\ref{fig_pos}), we
have scaled their observations to match the overall level of our new H
band observations.  For the south filament (Fig.~\ref{fig_comp_neb}a),
both spectra (S-NIR \& S-NIR(M97)) are very similar in the H band, but
show different K band continuum levels.  For the northwest filament
(Fig.~\ref{fig_comp_neb}b), the H band continuum is very similar
between NW-NIR and NW-NIR(M97), but both spectra have a much different
continuum shape than that of NW-NIR(M99-H2).  The K band spectra for
all three northwest filament spectra are similar in continuum shape
with NW-NIR(M97) having a slightly higher overall level.  The
differences between spectra are not surprising as the spectra from
different sources were taken with different apertures on different
positions in each filament.

As we have identified the origin of the northwest filament H band
continuum shape with a new dust emission feature (see
$\S$\ref{sec_newf}), it is important to point out that the H band
continuum shape is very unlikely to be the result of an error in the
way the data were taken or reduced.  If there was such an error, it
would be very unlikely that the H band spectral shapes of both the
south and northwest filaments would agree between our spectra (S-NIR
\& NW-NIR) and those of \citet{mar97} (S-NIR(M97) \& NW-NIR(M97)) as
each was taken by a different instrument on a different telescope by
different observers and reduced by different people.  Therefore, we
conclude that the spectral shape of the northwest H band spectra is
real.

\section{Discussion \label{sec_discuss}}

\begin{figure*}[tbp]
\epsscale{2.3}
\plottwo{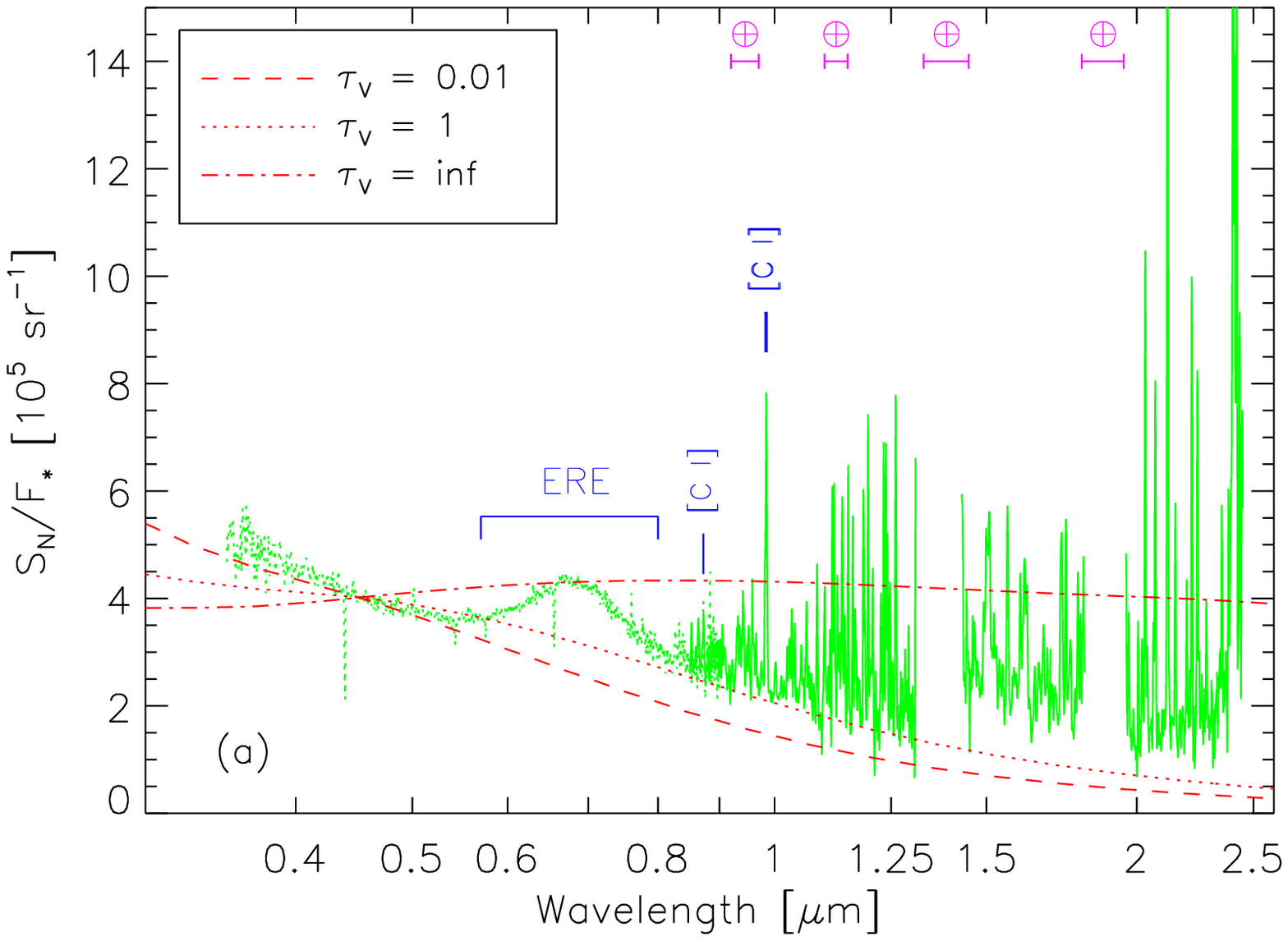}{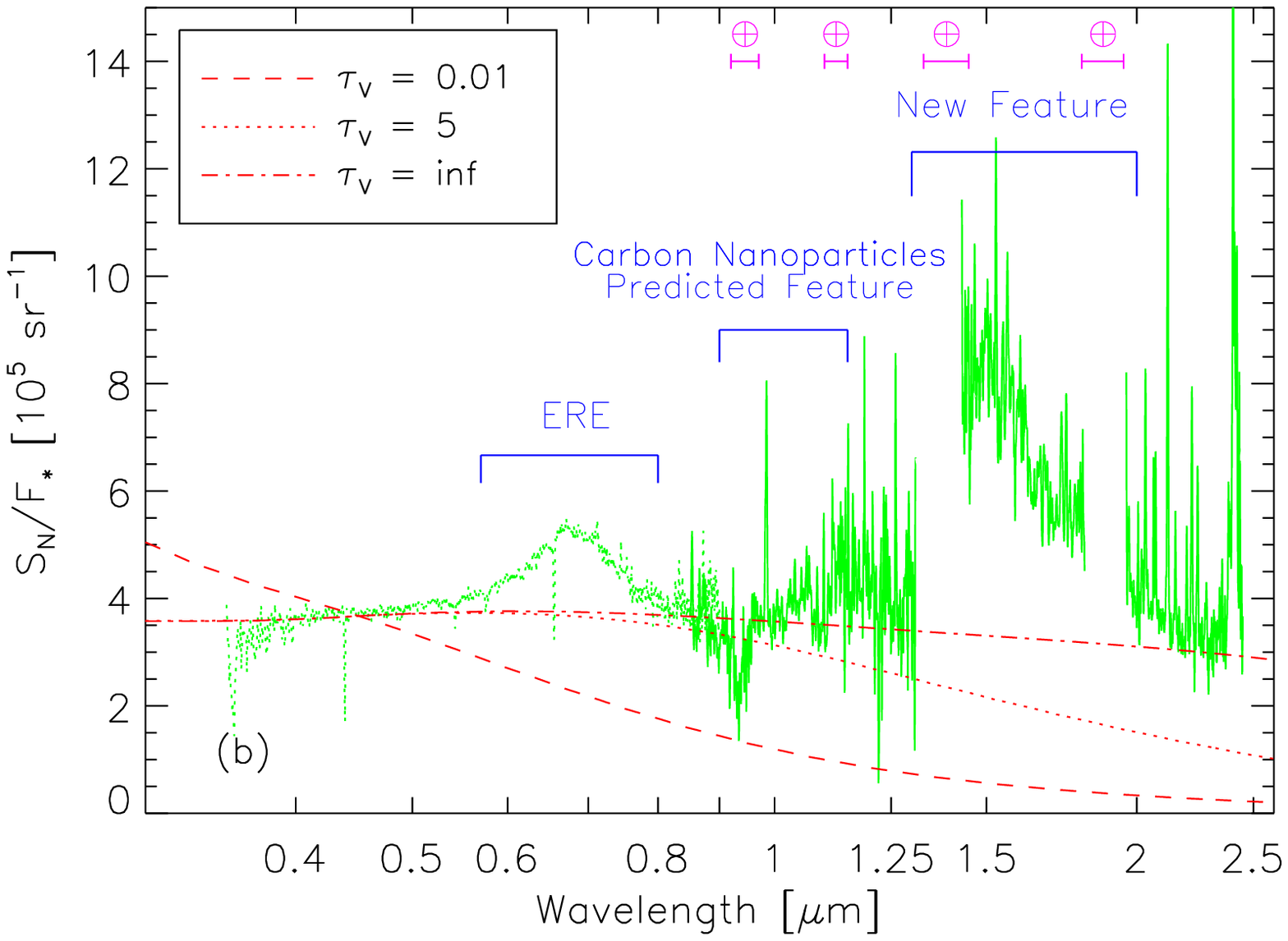}
\caption{The combined optical and near-infrared spectra for the (a)
south and (b) northwest filaments are shown.  The spectra are
displayed as the nebular surface brightness to stellar flux
($S_N/F_*$) ratio.  To compensate for offsets caused by the different
optical and near-infrared apertures used, the near-infrared spectra in
(a) were multiplied by 0.6 to match the optical data.  The noise-like
appearance of the near-infrared spectra is due to a virtual forest of
$H_2$ emission lines.  Almost all of the lines in the near-infrared
are due to $H_2$ with the notable exception of the two \ion{C}{1}
lines identified in (a).  The locations of the major atmospheric
absorptions in the near-infrared are shown at the top of both plots.
The smooth lines are theoretical ratio spectra computed as discussed
in \S3.1.
\label{fig_neb}} 
\end{figure*}

With the new spectra presented in this paper, we can study the dust
scattering and emission in NGC 7023 between 0.35 to 2.5 $\micron$.
This allows us to characterize the dust scattering and emission in the
optical and near-infrared and test the prediction of \citet{sea99}.
The 0.35 to 2.5 $\micron$ spectra of both filaments are plotted in
Figure~\ref{fig_neb}.  Both filament spectra have been divided by the
spectrum of HD 200775, leaving ratio spectra whose shape is only
dependent on the dust scattering, dust emission, and gas emission.
The gas emission gives rise the [\ion{C}{1}] and $H_2$ emission lines,
the analysis of which is beyond the scope of this paper.  The dust
scattering and emission mainly affects the continuum shape of these
ratio spectra while the gas emission is responsible for most of the
emission lines.

\subsection{Scattered Light}

In Fig.~\ref{fig_neb}, the theoretical ratio spectra for different
dust scattering situations have been plotted in addition to the
filament ratio spectra.  The theoretical spectra have been computed
using a model presented in \citet{wit85}.  The model equations can be
used to compute the ratio spectra for single angle dust scattering.
This is the situation for the two filaments of NGC 7023 which are
discrete density enhancements in the reflection nebula
\citep{ger98}.  The theoretical ratio spectra were computed using
average Milky Way dust grain properties (albedo, scattering phase
function asymmetry, and extinction curve) \citep{wol00}.  The cases
plotted in Fig.~\ref{fig_neb} are for infinite optical depth ($\tau_V
=$ inf), very optically thin ($\tau_V = 0.01$), and a best estimate
case ($\tau_V = 1$ for (a) and $\tau_V = 5$ for (b)).  We assumed the
optical depths from the star and the filament to the observer were
equal.  All the theoretical curves were normalized to the observed
value at 0.45~$\micron$.  The scattering angles which gave reasonable
fits to the data were 90$^{\circ}$ and 60$^{\circ}$ for the south
(Fig.~\ref{fig_neb}a) and northwest (Fig.~\ref{fig_neb}b) filaments,
respectively.  It was only possible to get {\em reasonable} fits to
the data as the continuum in the near-infrared is due to both dust
scattered stellar light and emission from transitionally heated small
dust particles
\citep{sel92}.  As a result, the fraction of the ratio spectrum due to 
dust scattering in the near-infrared is decreasing as the wavelength
increases.  The emission from transitionally heated small dust
particles starts in the J band and increases rapidly to the K band.

\subsection{Absence of 1 $\micron$ Carbon Nanoparticles Feature}

\citet{sea99} predicted that ERE is due to photoluminescence from
$sp^2$ coordinated carbon nanoparticles.  For the conditions present
in NGC 7023, they 
predicted that ERE produces 0.7 and 1.0~$\micron$ emission bands (see
their Fig.~4).  Both the observed filaments exhibit strong ERE bands near
0.7~$\micron$ as seen in Fig.~\ref{fig_neb} and previous work
\citep{wit88, wit90}.  Neither filament shows a 1.0~$\micron$
feature as predicted by \citet{sea99} for carbon nanoparticles.  The
northwest filament does show tentative evidence for a feature peaking
at 1.15 $\micron$ (see Fig.~\ref{fig_feature_id} and
\S\ref{sec_newf}), but this feature peaks at a significantly longer
wavelength than predicted by \citet{sea99} and so is unlikely to be
due to carbon nanoparticles.  The central wavelength of the
1.0~$\micron$ peak predicted by \citet{sea99} varies very little ($<
0.02$~$\micron$) for a large range of carbon particle sizes
($\left<N_C\right> = 
125 - 500$) and dispersions around each size ($\sigma = 125 - 500$).
The lack of any feature at 1.0~$\micron$ in both filaments calls into
question the identification of carbon nanoparticles as modeled by
\citet{sea99} as the carrier of
ERE.  We do note that \citet{sea99} do predict a weak 1.0~$\micron$
peak for a small subset ($\left<N_C\right> \sim 125 - 200$ and $\sigma
\sim 125 - 
150$) of the size distributions of carbon nanoparticles they modeled
(see their Figs.~2 \& 4) and state in their conclusions that there are
mixtures with larger carbon nanoparticles which exhibit the main ERE
peak around 0.8~$\micron$ and no 1.0~$\micron$ peak.  But this does
not correspond to the conditions in the NGC~7023 filaments which have
their main ERE peaks near 0.7~$\micron$.

While the lack of the 1.0~$\micron$ peak in our observations does not
rule out carbon nanoparticles as the source of ERE, it does diminish
the probability that they are the source of ERE.  This is especially
true when the minimum 10\% photoluminescense efficiency of ERE in the
diffuse ISM \citep{gor98} is considered.  \citet{sea99} chose to only
model carbon nanoparticles with $sp^2$ bonds specifically so that the
particles would have a high photoluminescence efficiency.  Carbon
nanoparticles with $sp^2$ and $sp^3$ bonds of different sizes have
been produced in the lab \citep{her98} and have a negligible
photoluminescence efficiency (N.\ Herlin, private communication).
Thus, the main problem with carbon nanoparticles being responsible for
ERE is their low efficiency as measured in the laboratory.  The
\citet{sea99} carbon nanoparticles model can account for this, but the 
lack of a 1.0~$\micron$ peak in NGC 7023 does not confirm their model.
It is likely the \cite{sea99} model can be modified to account for our
new observations (W.\ Duley, private communication), but their model
needs to make predictions which are confirmed by observations or be
directly supported by laboratory work before carbon nanoparticles can
be convincingly identified with ERE.  As such it is worth noting that
silicon nanoparticles do a better job of reproducing the observed
characteristics of ERE and their characteristics are supported by
laboratory measurements \citep{led98, wit98, cre99, led00}.  We have
found tentative additional evidence for silicon nanoparticles in our
observations which is discussed below.

\subsection{Two New Dust Emission Features \label{sec_newf}} 

\begin{figure*}[tbp]
\epsscale{1.5}
\plotone{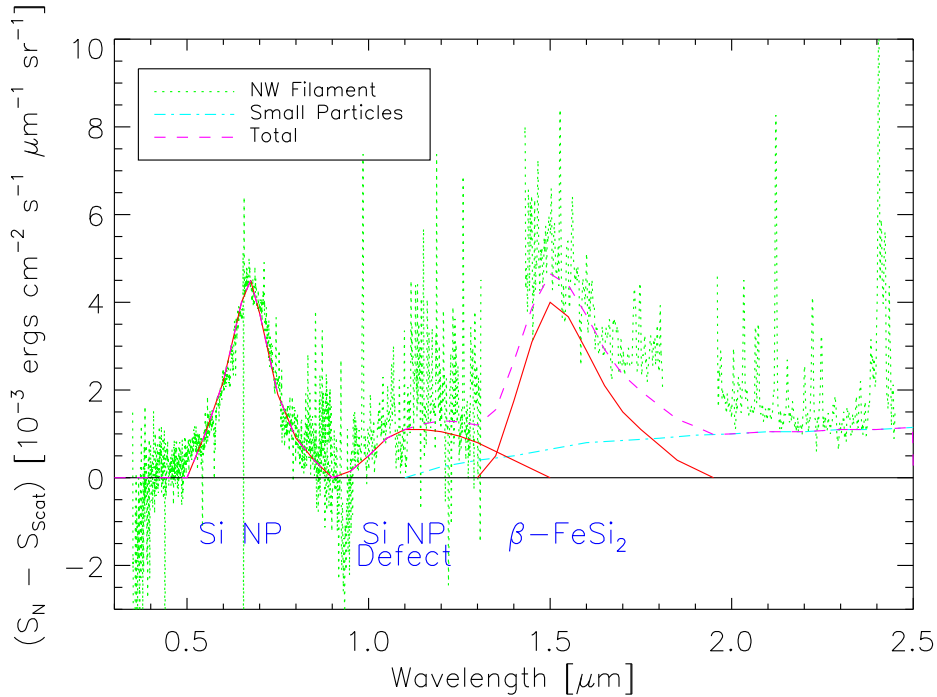}
\caption{The northwest filament emission spectrum (observed minus
scattered light model) is plotted (dotted line).  In addition,
profiles of the proposed carriers (solid lines), emission from
transitionally heated small particles (dash-dot line), and the total
(dashed line) are plotted.  The 0.7 $\micron$ silicon nanoparticle
photoluminescence band is from \citet{guh97} and
\citet{ehb97}.  The 1.1 $\micron$ silicon nanoparticles defect band is 
from \citet{fuj98, fuj99}.  The 1.5 $\micron$ $\beta$-FeSi$_2$ band is
from \citet{leo96} and \citet{sue99}. \label{fig_feature_id}}
\end{figure*}

The unexpected result of this study is the evidence for at least one,
and possibly two, previously unobserved, broad emission features which 
peak in the near-infrared.  Examining our spectra of NGC 7023 and
those in the literature \citep{mar97,mar99}, the new features are seen
in only part of the northwest filament (see Figs.~\ref{fig_pos} and
\ref{fig_comp_neb}).  In order to show the profiles and relative
strengths of the new features as well as the ERE band, we display the
spectrum of the intensity of the northwest filament, with the
scattering spectrum for $\tau_v = 5$ (Fig.~\ref{fig_neb}b) subtracted
in Fig.~\ref{fig_feature_id}.  This spectrum is due to excess {\em
emission} arising from the dust and gas in the northwest filament.
Here we will focus solely on the dust emission components.

We identify three discrete emission features, in addition to a broad
continuum which increases in intensity toward longer wavelengths. This
broad continuum has been previously identified as non-equilibrium
thermal emission from stochastically heated tiny particles in the
sub-nm size range \citep{sel92}. The discrete features in the
northwest filament have peaks at 0.68, 1.15, and 1.50~$\micron$, and
widths characteristic of solid-state photoluminescence bands.  The
0.68 band is the well known ERE band.

The strongest of the two new bands is at 1.50~$\micron$ and is similar
in strength and width to the visible ERE band at 0.68~$\micron$. No
interstellar band with these characteristics has ever been reported
before in this wavelength region. Given the strength of this band, an
origin in a photoluminescence process involving solids made from
heavily depleted abundant refractory elements is strongly suggested.
The only band we found, after an extensive search of the astronomy and
physics literature, fitting this description is the photoluminescence
band in the low-temperature, amorphous phase of iron disilicide
($\beta$-FeSi$_2$) \citep{leo96, sue98, sue99, fil99}.  Iron
disilicide is a direct-bandgap semiconductor with a bandgap near 0.8
eV, in which cross-bandgap recombination of free electrons and holes
produces photoluminescence similar to that observed in silicon
nanoparticles. However, in contrast to the case in silicon, an
indirect-bandgap semiconductor, FeSi$_2$ does not need to be in the
form of nanoparticles in order to exhibit efficient
photoluminescence. In fact, low-temperature FeSi$_2$ particles of
about 100~nm diameter appear to be luminescing most efficiently, with
the width of the band dependent upon the state of annealing, with
relatively narrow bands seen only with annealing temperatures in
excess of 900$\arcdeg$~C. These temperatures are not likely to
occur in large 
dust grains located in the NGC 7023 filaments as the radiation field
density is unlikely to heat the dust grains to such high
temperatures.  Thus, both the width and relative
strength of the band at 1.5~$\micron$ are consistent with
photoluminescence by low-temperature iron disilicide grains.  In
Fig.~\ref{fig_feature_id}, we have plotted the laboratory FeSi$_2$
band profile which most closely fits the observed 1.5~$\micron$
feature.  The observed profile does deviate from the laboratory
profile on the long wavelength side.  This deviation may be due to the
fact that the laboratory profile is for a single particle size and the
observed profile is likely the result of a population of different
sized particles.  We have no explanation at present why such grains
are observable in the northwest filament and not in the south
filament. More observations of this interesting new band in other
reflection nebulae and under a wider range of physical conditions are
clearly needed.

Our tentative identification of the new 1.5~$\micron$ feature with
FeSi$_2$ dust 
grains is supported by two recent papers \citep{fer00, hen00}.
\citet{fer00} performed 
chemical equilibrium calculations for a particular gas phase mixture
deficient in carbon and oxygen 
and found that solid FeSi is the first condensate of abundant
refractory elements.  They identify the
previously unidentified, strong 47.5~$\micron$ feature in the spectrum
of the evolved stars AFGL 4106 \citep{mol99} as arising from FeSi
grains.  It is possible that FeSi$_2$ could also form in similar
mixtures of refractory elements (H.-P.\ Gail, private communication).
\citet{hen00} presents a laboratory spectrum of FeSi$_2$ grains from
10 to 70~$\micron$.  The main features of this spectrum are bands at
21, 27, and 30 $\micron$, and similar bands have been seen in IRAS and
ISO data \citep{hen00}.  Combining these two papers and our work imply
that FeSi$_2$ dust grains could very well be present in the
interstellar medium. 
Additional observations are obviously needed, especially aperture
matched observations from 1 to 35~$\micron$.

The weaker of the two new bands peaks around 1.15~$\micron$ and
its 
presence in the northwest filament spectrum is tentative at best.
Additional observations are definitely required to confirm the
presence of this feature.  It is intriguing to note the wavelength
location of the new 1.15~$\micron$ feature in light of recent
laboratory work on silicon nanoparticles.  Laboratory studies of
photoluminescence of silicon nanoparticles in the form of porous
silicon or matrix-embedded nano-crystallites \citep{mey93, gar94,
pet95, hil96a, hil96b, fuj98, fuj99} have produced evidence for the
existence of a second, near-IR, photoluminescence band, in addition to
the main optical band matching the ERE.  The peak wavelengths of these
two bands, both subject to quantum confinement, vary with the size of
the emitting particles and are closely correlated by the relation
\begin{equation}
E_p(IR) = 0.43 E_p(ERE) + 0.34~~[\mbox{in eV}],
\label{eq_sinp}
\end{equation}
where $E_p(IR)$ and $E_p(ERE)$ are the peak energies of the near-IR
band and the visible band in eV, respectively \citep{hil96b}. With the
ERE observed to peak at 0.68~$\micron$ in the northwest filament,
Eq.~\ref{eq_sinp} predicts a peak wavelength for the near-IR band of
1.1~$\micron$. We have indicated the silicon nanoparticle bands in
accordance with laboratory results in Fig.~\ref{fig_feature_id}. They
provide very satisfactory fits to the observational data.

The principal ERE band in the visible is attributed to recombination
of free electron-hole pairs in silicon nanoparticles (Si NP in
Fig.~\ref{fig_feature_id}) \citep{wit98}, while the near-IR band from
the same type of particles is attributed to recombination of photoexcited
electrons from the conduction band to in-bandgap defect sites
associated with Si dangling bonds at the surfaces of silicon
nano-crystallites \citep{fuj98}. The near-IR band is observed in the
laboratory only at temperatures of $T < 150 K$, while the visible band
remains strong at higher temperatures.  The relative strength of the
two bands also depends on the degree of passivation of surface
dangling bonds. With increasing passivation (e.g., with oxygen atoms)
the near-IR band decreases in strength while the visible band gains in
strength.  The fact that only the northwest filament shows the near-IR
band near 1.15~$\micron$ but both the south and the northwest
filaments display the visible ERE band is thus easily explained.  The
maximum temperature reached by individual nanoparticles in the south
filament upon absorption of individual photons is either higher than
the corresponding temperature of nanoparticles in the northwest
filament, or the degree of passivation is higher in the south
filament. The first explanation seems more likely in view of the
higher optical depth of the northwest filament ($\tau_V \sim 5$)
compared to the more optically thin south filament ($\tau_V \sim
1$). Energetic photons are thus less likely to reach nanoparticles in
the northwest filament. \citet{wit98} estimated that a 1.7~nm diameter
Si nanoparticle experiences a temperature increase of $\sim$170 K upon
absorbing a 10~eV photon. We conclude that the average maximum
temperature of the Si nanoparticles in the northwest filament must be
lower than those in the south filament, low enough to allow the
near-IR band to be observable.

Given that the visible ERE band has been observed with peak
wavelengths ranging from 0.65 to 0.88~$\micron$ in different
environments, Eq.~\ref{eq_sinp} predicts the possible existence of
corresponding near-IR bands with peak wavelengths spanning the range
from 1.07 to 1.31~$\micron$. Detection of such bands with wavelengths
fitting the correlation in Eq.~\ref{eq_sinp} would be a powerful
confirmation of the silicon nanoparticle model for the ERE. Additional
observations in the near-IR region such as reported here are clearly
needed.

\section{Conclusions}

From new 0.35 to 2.5 $\micron$ spectra of the south and northwest
filaments in the reflection nebula NGC 7023, we conclude that:
\begin{itemize}
\item The 1.0 $\micron$ feature predicted by \citet{sea99} is not seen.
This result when combined with the negligible measured
photoluminescence efficiency of carbon nanoparticles calls into
question the identification of carbon nanoparticles as the source for
ERE.
\item We have discovered a new strong and broad dust emission feature
which peaks at 1.5~$\micron$, and we tentatively identify it with
photoluminescence from $\beta$-FeSi$_2$ grains.
\item We have also found tentative evidence for another new broad,
dust emission feature at 1.15~$\micron$.  We identify this feature
with photoluminescence from transitions from the conduction band to
defect states in silicon nanoparticles.  If this is correct, it lends
support to the identification of silicon nanoparticles as the carrier
of ERE.
\end{itemize}

\acknowledgements

This work has benefited from conversations with C.\ Engelbracht, M.\
Rieke, and G.\ Rieke.  We thank P.\ Martini for providing us with
electronic versions of his NGC 7023 near-IR spectroscopy.  A.\ N.\
Witt acknowledges fruitful exchanges with M.\ Fujii concerning the
identification of the two new infrared features.  We gratefully
acknowledge the constructive comments of the referee, Th.\ Henning.
A.\ N.\ Witt and T.\ L.\ Smith acknowledge support from NAG5-4338 to
the Univ.\ of Toledo.  R.\ Rudy, D.\ Lynch, \& S.\ Mazuk acknowledge
support from The Aerospace Corporation's Independent Research and
Development Program directed to the Space Science Applications
Laboratory.

\begin{deluxetable}{lllccc}
\tablewidth{0pt}
\tablecaption{Filament Observations \label{tab_pos}}
\tablehead{\colhead{name} & \colhead{wavelength} & 
    \colhead{position\tablenotemark{a}} 
   & \colhead{aperture} & \colhead{position angle} & \colhead{ref} \\
    & \colhead{[$\micron$]} & \colhead{[$\arcsec$]} & 
     \colhead{[$\arcsec$]} & \colhead{[$^{\circ}$]} & }
\startdata
\multicolumn{6}{c}{Northwest Filament} \\ \tableline
NW-OPT & 0.35--0.9 & 14.5 W, 52 N & $3 \times 21.1$ & 5.7 & 1 \\
NW-NIR & 0.85--2.5 & 46 W, 26 N & $3 \times 45$ & 60 & 1 \\
NW-NIR(M97) & 1.5--2.5 & 40 W, 34 N & $0.6 \times 5$ & 0 & 2 \\
NW-NIR(M99-N) & 1.2--2.5 & 27 W, 48.5 N & $4.5 \times 9$ & 0 & 3 \\
NW-NIR(M99-H2) & 1.2--2.5 & 27 W, 36.5 N & $4.5 \times 9$ & 0 & 3 \\
NW-NIR(M99-S) & 1.2--2.5 & 27 W, 22 N & $4.5 \times 9$ & 0 & 3 \\ \tableline
\multicolumn{6}{c}{South Filament} \\ \tableline
S-OPT & 0.35--0.9 & 27 W, 71 S & $3 \times 13.3$ & 5.7 & 1 \\
S-NIR & 0.85--2.5 & 11 W, 66 N & $3 \times 45$ & 90 & 1 \\
S-NIR(M97) & 1.5--2.5 & 22 W, 66 S & $0.6 \times 5$ & 0 & 2 \\
\enddata
\tablenotetext{a}{Positions are referenced to HD 200775 ($21^h 01^m
36\fs 92$, $+68\arcdeg 09\arcsec 47\farcs 763$, 2000.0), the central
star of NGC 7023.}
\tablerefs{(1) this work; (2) \citet{mar97}; (3) \citet{mar99}}
\end{deluxetable}


\begin{thebibliography}{}
\bibitem[Allen (1973)]{all73} Allen, D.\ A.\ 1973, \mnras, 161, 145
\bibitem[Cesarsky et al.\ (1996)]{ces96} Cesarsky, D., Lequeux, J.,
   Abergel, A., Perault, M., Palazzi, E., Madden, S., \& Tran, D.\
   1996, \aap, 315, L305
\bibitem[Credo, Mason, \& Buratto (1999)]{cre99} Credo, G.\ M.,
   Mason, M.\ D., \& Buratto, S.\ K.\ 1999, Appl.\ Phys.\ Lett., 74,
   1978 
\bibitem[Duley (1985)]{dul85} Duley, W.\ W.\ 1985, \mnras, 215, 259
\bibitem[Duley \& Seahra (1999)]{dul99} Duley, W.\ W., \& Seahra, S.\
   S.\ 1999, \apj, 522, L129
\bibitem[Ehbrecht et al.\ (1997)]{ehb97} Ehbrecht, M., Kohn, B.,
   Huisken, F., Laguna, M.\ A., \& Paillard, V.\ 1997, \prb, 56(11),
   6958 
\bibitem[Ferrarotti et al.\ (2000)]{fer00} Ferrarotti, A., Gail,
   H.-P., Degiorgi, L., \& Ott, H.\ R.\ 2000, \aap, 357, L13
\bibitem[Filonov et al.\ (1999)]{fil99} Filonov, A.\ B., Borisenko,
   V.\ E., Henrion, W., \& Lange, H.\ 1999, J.\ of Luminescence, 80,
   479 
\bibitem[Fuente et al.\ (2000)]{fue00} Fuente, A., Mart\'{i}n-Pintado, 
   J., Rodr\'{i}guez-Fern\'{a}ndez, N.\ J., Cernicharo, J.\ \& Gerin,
   M.\ 2000, \aap, 354, 1053
\bibitem[Fujii, Hayashi, \& Yamamoto (1998)]{fuj98} Fujii, M.,
   Hayashi, S., \& Yamamoto, K.\ 1998, Rec.\ Res.\ Devel.\ in Appl.\
   Phys., 1, 193 
\bibitem[Fujii et al.\ (1999)]{fuj99} Fujii, M., Mimura,
   A., Hayashi, S., \& Yamamoto, K.\ 1999, Appl.\ Phys.\ Lett.,
   75(2), 193 
\bibitem[Gardelis \& Hamilton (1994)]{gar94} Gardelis, S., \&
   Hamilton, B.\ 1994, J.\ Appl.\ Phys., 76, 5327 
\bibitem[Gerin et al.\ (1998)]{ger98} Gerin, M., Phillips, T.\ G.,
   Keene, J., Betz, A.\ L., \& Boreiko, R.\ T.\ 1998, \apj, 500, 329
\bibitem[Gordon, Witt, \& Friedmann (1998)]{gor98} Gordon, K.\ D., Witt,
   A.\ N., \& Friedmann, B.\ C.\ 1998, ApJ, 498, 522
\bibitem[Guha (1997)]{guh97} Guha, S.\ 1997, Thin Solid Films, 297, 102
\bibitem[Henning (2000)]{hen00} Henning, T.\ 2000, in ``ISO beyond 
   the peaks: The 2nd ISO workshop on analytical spectroscopy'', (ESA
   SP-456), in press
\bibitem[Herbst \& Shevchenko (1999)]{her99} Herbst, W., \&
   Shevchenko, V.\ S.\ 1999, \aj, 118, 1043
\bibitem[Herlin et al.\ (1998)]{her98} Herlin, N., Bohn, I., Reynaud,
   C., Cauchetier, M., Aymeric, G., \& Rouzaud, J.-N.\ 1998, \aap,
   330, 1127
\bibitem[Hill \& Whaley (1996a)]{hil96a} Hill, N.\ A., \& Whaley, K.\
   B.\ 1996a, J.\ Electronic Mat., 25, 269 
\bibitem[Hill \& Whaley (1996b)]{hil96b} Hill, N.\ A., \& Whaley, K.\
   B.\ 1996b, \prl, 76(16), 3039 
\bibitem[Hoffleit \& Jaschek (1982)]{hof82} Hoffleit, D., \& Jaschek,
   C.\ 1982, The Bright Star Catalog, (New Haven, Connecticut:  
   Yale University Observatory)
\bibitem[Koornneef (1983)]{koo83} Koornneef, J.\ 1983, \aap, 128, 84
\bibitem[Kurucz (1991)]{kur91} Kurucz, R.\ L.\ 1991, Precision Astronomy
   and Astrophysics of the Galaxy, ed.\ A.\ G.\ Davis  
\bibitem[Ledoux et al.\ (1998)]{led98} Ledoux, G., et al.\ 1998, \aap, 
   333, L39
\bibitem[Ledoux et al.\ (2000)]{led00} Ledoux, G., Guillois, O.,
   Reynaud, C., Huiskin, F., Kohn, B., \& Paillard, V.\ 2000, Mat.\
   Sci.\ \& Eng., B69-70, 350
\bibitem[Lemaire et al.\ (1996)]{lem96} Lemaire, J.\ L., Field, D.,
   Gerin, M., Leach, S., Peneau des For\^{e}ts, G., Rostas, F.\ \&
   Rouan, D.\ 1996, \aap, 308, 895
\bibitem[Lemaire et al.\ (1999)]{lem99} Lemaire, J.\ L., Field, D.,
   Maillard, J.\ P., Peneau des For\^{e}ts, G., Falgarone, E.,
   Pijpers, F.\ P., Gerin, M.\ \& Rostas, F.\ 1999, \aap, 349, 253
\bibitem[Leong et al.\ (1996)]{leo96} Leong, D.\ N., Harry, M.\ A.,
   Reeson, K.\ J., \& Homewood, K.\ P.\ 1996, Appl.\ Phys.\ Lett.,
   68(2), 1649 
\bibitem[Martini, Sellgren, \& Hora (1997)]{mar97} Martini, P.,
   Sellgren, K., \& Hora, J.\ L.\ 1997, \apj, 484, 296
\bibitem[Martini, Sellgren, \& DePoy (1999)]{mar99} Martini, P.,
   Sellgren, K., \& DePoy, D.\ L.\ 1999, \apj, 526, 772
\bibitem[Meyer et al.\ (1993)]{mey93} Meyer, B.\ K., Hofmann, D.\ M.,
   Stadler, W., Petrova-Koch, V., Koch, F., Omling, P., \&
   Emanuelsson, P.\  1993, Appl.\ Phys.\ Lett., 63, 2120 
\bibitem[Misselt, Clayton, \& Gordon (1999)]{mis99} Misselt, K.\ A.,
   Clayton, G.\ C., \& Gordon, K.\ D.\ 1999, \pasp, 111, 1398
\bibitem[Molster et al.\ (1999)]{mol99} Molster, F.\ J., et al.\ 1999, 
   \aap, 350, 163
\bibitem[Moutou et al.\ (1999)]{mou99} Moutou, C., Verstraete, L.,
   Sellgren, K. \& Leger, A.\ 1999, in ``The Universe as seen by ISO'',
   eds.\ P.\ Cox, V.\ Demuyt, \& M.\ Kessler, ESA SP-427, 727 
\bibitem[Papoular et al.\ (1996)]{pap96} Papoular, R., Conard, J.,
   Guillois, O., Nenner, I., Reynaud, C., \& Rouzaud, J.-N.\ 1996,
   \aap, 315, 222
\bibitem[Petrova-Koch \& Muschik (1995)]{pet95} Petrova-Koch, V.,
   \& Muschik 1995, Thin Solid Films, 255, 246 
\bibitem[Rudy, Puetter, \& Mazuk (1999)]{rud99} Rudy, R.\ J., Puetter,
   R.\ C., \& Mazuk, S.\ 1999, \aj, 118, 666
\bibitem[Sakata et al.\ (1992)]{sak92} Sakata, A., Wada, S., Narisawa, 
   T., Asano, Y., Iijima, Y., Onaka, T., \& Tokunagu, A.\ T.\ 1992,
   \apj, 393, L83
\bibitem[Seahra \& Duley (1999)]{sea99} Seahra, S.\ S., \& Duley, W.\
   W.\ 1999, \apj, 520, 719
\bibitem[Sellgren (1983)]{sel83} Sellgren, K.\ 1983, \aj, 88, 985
\bibitem[Sellgren et al.\ (1985)]{sel85} Sellgren, K., Allamandola,
   L.\ J., Bregman, J.\ D., Werner, M.\ W.\ \& Wooden, D.\ H.\ 1985,
   \apj, 299, 416
\bibitem[Sellgren, Werner, \& Allamondola (1996)]{sel96} Sellgren, K.,
   Werner, M.\ W., \& Allamandola, L.\ J.\ 1996, \apjs, 102, 369
\bibitem[Sellgren, Werner, \& Dinerstein (1992)]{sel92} Sellgren, K.,
   Werner, M.\ W., \& Dinerstein, H.\ L.\ 1992, \apj, 400, 238
\bibitem[Schnaiter et al.\ (1998)]{sch98} Schnaiter, M., Mutschke, H., 
   Dorschner, J., Henning, Th., \& Salama, F.\ 1998, \apj, 498, 486
\bibitem[Schnaiter et al.\ (1999)]{sch99} Schnaiter, M., Henning, Th., 
   Mutschke, H., Kohn, B., Ehbrecht, M., \& Huisken, F.\ 1999, \apj,
   519, 687
\bibitem[Stapelfeldt, K.\ et al.\ (1997)]{sta97} Stapelfeldt, K.\ et
   al.\ 1997, in ``Planets Beyond the Solar System and the Next
   Generation of Space Missions'', ed.\ D. Soderblom, ASP Conference
   Series vol.\ 119, 131
\bibitem[Suemasu et al.\ (1998)]{sue98} Suemasu, T.,  
   Fujii, T., Iikura, Y., Takakura, K., \& Hasegawa, F.\ 1998,
   Jpn.\ J.\ of Appl.\ Phys., 37, L1513 
\bibitem[Suemasu et al.\ (1999)]{sue99} Suemasu, T., Iikura, Y.,
   Fujii, T., Takakura, K., Hiroi, N., \& Hasegawa, F.\ 1999, Jpn.\
   J.\ Appl.\ Phys., 38, L620  
\bibitem[Szomoru \& Guhathakurta (1998)]{szo98} Szomoru, A., \&
   Guhathakurta, P.\ 1998, \apj, 494, L93
\bibitem[Uchida, et al.\ 2000]{uch00} Uchida, K.\ I., Selgren, K.,
   Werner, M.\ W., \& Houdashelt, M.\ L.\ 2000, \apj, 530, 817
\bibitem[Webster (1993)]{web93} Webster, A.\ 1993, \mnras, 264, L1
\bibitem[Witt (1985)]{wit85} Witt, A.\ N.\ 1985, \apj, 216, 224
\bibitem[Witt \& Boroson (1990)]{wit90} Witt, A.\ N., \& Boroson, T.\
   A.\ 1990, \apj, 355, 182 
\bibitem[Witt \& Cottrell (1980)]{wit80} Witt, A.\ N., \& Cottrell,
   M.\ J.\ 1980, \aj, 85, 22
\bibitem[Witt, Gordon, \& Furton (1998)]{wit98} Witt, A.\ N., Gordon,
   K.\ D., \& Furton, D.\ G.\ 1998, \apj, 501, L111
\bibitem[Witt et al.\ (1992)]{wit92} Witt, A.\ N., Petersohn, J.\ K.,
   Bohlin, R.\ C., O'Connell, R.\ W., Roberts, M.\ S., Smith, A.\ M.,
   \& Stecher, T.\ P.\ 1992, \apj, 395, L5
\bibitem[Witt \& Schild (1988)]{wit88} Witt, A.\ N., \& Schild, R.\
   E.\ 1988, \apjs, 62, 839
\bibitem[Wolff, Clayton, \& Gordon (2000)]{wol00} Wolff, M.\ W.,
   Clayton, G.\ C., \& Gordon, K.\ D.\ 2000, in preparation
\end{thebibliography}
\end{document}